\documentclass{article}

\usepackage[preprint]{neurips_2019}

\usepackage{hyperref}       % hyperlinks
\usepackage{url}            % simple URL typesetting
\usepackage{amsfonts}       % blackboard math symbols
\usepackage{graphicx}
\usepackage{eqnarray,amsmath}
\usepackage[colorinlistoftodos]{todonotes}
\usepackage{multirow}
\usepackage{verbatim}

\title{Task Driven Sensor Fusion: A Perspective On Learning Interpretable Multiagent Interactions}
\title{Learning Interpretable Multiagent Communication: A Case Study}
\title{Learning Interpretable Multiagent Communication: A Numerical Case Study}
\title{Learning Interpretable Multiagent Communication for Multimodal Information Fusion}
\title{A perspective on multi-agent communication for information fusion}
\author{%
  Homagni Saha\\
  Department of Mechanical Engineering\\
  Iowa State University\\
  Ames, IA 50011 \\
  \texttt{hsaha@iastate.edu} \\
  \And
  Vijay Venkataraman \\
  Honeywell Aerospace \\
  Plymouth, MN 55441 \\
  \texttt{Vijay.Venkataraman@honeywell.com} \\
  \AND
  Alberto Speranzon \\
  Honeywell Aerospace \\
  Plymouth, MN 55441 \\
  \texttt{Alberto.Speranzon@honeywell.com} \\
  \And
  Soumik Sarkar \\
  Department of Mechanical Engineering\\
  Iowa State University\\
  Ames, IA 50011 \\
  \texttt{soumiks@iastate.edu} \\
}

\begin{document}
\maketitle
\vspace*{-0.3cm}
\begin{abstract}
  Collaborative decision making in multi-agent systems typically requires a predefined communication protocol among agents. Usually, agent-level observations are locally processed and information is exchanged using the predefined protocol, enabling the team to perform more efficiently than each agent operating in isolation. In this work, we consider the situation where agents, with complementary sensing modalities must co-operate to achieve a common goal/task by learning an efficient communication protocol. We frame the problem within an actor-critic scheme, where the agents learn optimal policies in a centralized fashion, while taking action in a distributed manner. We provide an interpretation of the emergent communication between the agents. We observe that the information exchanged is not just an encoding of the raw sensor data but is, rather, a specific set of directive actions that depend on the overall task. Simulation results demonstrate the interpretability of the learnt communication in a variety of tasks.
 \end{abstract}
  %Information fusion from multiple sensor sources is an essential ingredient in the perception and decision making systems. Traditionally, the problem is approached in ways that consider developing joint feature representations suitable for the task at hand \toso{add some relevant references here}. We consider a different approach where a multiagent system comprising of different agents each specializing in sensing unique modalities may fuse individual information through learned communication protocols. We test our idea in a variety of navigation scenarios and find emergence of human interpretable communication strategies.
%\todo[inline]{There is the need to make the wording consistent: we use environment/task/scenario, we say color/shape agent and color-/shape-agent, multimodal and complementary sensing, etc. i fixed a few but need to go from top to bottom to make it consistent. There need to be a bit more discussion on related work. If we run out of space we could combine subsections 3.3 and 3.4: the headings have lot of space around. Conclusions should be a bit ``beefier''.}
\section{Introduction}

In this paper, we analyze communication protocols learnt by a team of agents equipped with complementary sensor modalities and tasked with a common goal.  We call this ``task based multi-modal decision making'', wherein agents learn to map their sensor measurements and the information communicated by other agents, directly into actions based on the common goal. In this setting, each agent has access only to its own sensor data but needs to rely on the communication with other agents to obtain task relevant information from that agent's sensor modality. We present a way to interpret the emergent communication by visualizing this mapping into the agent's action space. We find the communication that is learnt, within a reinforcement learning paradigm, is not only emergent, but is task dependent and adaptive to the size of the communication channel.  

In relation to existing literature on learning for multi-agent systems our work borrows from the general framework of ``Markov-games'', proposed in~\cite{lowe2017multi,mordatch2018emergence,foerster2016learning} and references therein. In particular, we consider (~\cite{lowe2017multi}) for our learning problem. Related to the emergent communication aspect, central to this work, we consider ideas from the literature on ``multi-agent referential games'' (~\cite{golland2010game,andreas2016reasoning,evtimova2017emergent,lazaridou2018emergence}), where a (sender) agent communicates highly structured information (images and text) to a (receiver) agent which has to interpret what the other agent saw. However, here we are interested in the evolution of communication for unstructured data under joint interactions using an actor-critic algorithm. Emergent communication was also studied in (~\cite{kottur2017natural,cao2018emergent}), however, we focus our analysis of emergent communication on the action space. Specifically, we project the learnt communication on the action space and visually analyze the results. This enables us to more clearly interpret the learnt communication. It is shown that powerful joint representations of the world can be encoded through task dependent communication which is easy to interpret under complementary sensing modality constraints.

\section{Environment and tasks} 
%Each agent's state is continuous (location), whereas we assume the actions to be discrete. Each agent's observation is defined as a vector $o_i = [x_{i1},m_{i1},\dots,x_{ij},m_{ij},c_{i1},\dots,c_{ik},g_{i1},\dots,g_{il}]$, where $x_{ij}$ denotes the horizontal and vertical distances from the $i^{th}$ agent to the $j^{th}$ landmark; $m_{ij}$ denotes an encoding of the property of the $j^{th}$ landmark detected by the $i^{th}$ agent. The vectors $w_{1}=[1,0,0]$, $w_{2}=[0,1,0]$, $w_{3}=[0,0,1]$ denote that a landmark has the color red, green or blue, respectively, when observed by the color-agent. We use the same vectors to denote that a landmark has the shape circle, square and triangle when observed by the shape-agent. The vector $c_{i1},\dots,c_{ik}$, part of the agent $i^{th}$ action space, denotes a~$k$ bit stream. 
For our experimental study, we consider a two-dimensional world with two agents and $L$ landmarks. Each landmark has a color $\{red, green, blue\}$ and shape $\{triangle, circle, square\}$ property. Our agents have complementary sensing modalities: one of the agents, denoted as color-agent, can only observe the color of the landmarks and the other, denoted as shape-agent, can only observe the shape of the landmarks. We assume that both agents can measure their (relative) distance from all landmarks but cannot measure their distance from each other. At every discrete time step the agents take both physical movement actions (a unit movement in one of the four directions or stand still) and communicative actions, namely, broadcast a $k$-bit message. The communication message sent by one agent is received by the other in the next time step.\\ %Each agent's state is continuous (location), whereas we assume the actions to be discrete. 
Each agent's observation is a vector $\mathbf{o_i} = [x_{i1},y_{i1},\mathbf{m_{i1}},\dots,x_{iL},y_{iL},\mathbf{m_{iL}} | c_{i1},\dots,c_{ik}|\allowbreak g_{i1},\dots,g_{iL}]$, where $x_{ij}$ and $y_{ij}$ denote the horizontal and vertical distances, respectively, between the $i^{th}$ agent and the $j^{th}$ landmark; $\mathbf{m_{ij}}$ denotes a one hot encoding of the $j^{th}$ landmark's property as sensed by the $i^{th}$ agent. For example, $\mathbf{m_{1j}} \in \{ {[1,0,0], [0,1,0], [0,0,1]}\}$ denotes the encoding of the $j^{th}$ landmark's color for the color agent or shape for the shape agent. The vector $[c_{i1},\dots,c_{ik}]$, denotes the~$k$ bit word received bythe $i^{th}$ agent. Finally, $[g_{i1},\dots,g_{iL}]$ denotes  one hot encoding of the target landmark properties provided to agent~$i$. We base our study on the three collaborative tasks, described below. \\
\textbf{Task 1: Cross modal information exchange:} In this task, the map contains three landmarks. No two landmarks have the same shape or color. During each episode, agents and landmarks are placed randomly in the map. One of the landmarks is designated as ``target'' and the goal, for both agents, is to reach the designated target landmark. This target landmark's property is indicated to the agents using $[g_{i1},\dots,g_{iL}]$ as described previously. Consider as example where the target landmark is a blue circle, if we were to provide to the color agent the color properties, it can trivially navigate to the target given that it has full knowledge of where the different colored landmarks are with respect to itself. To avoid this and to encourage communication, we pass the encoding corresponding to the shape of the target landmark to the color-agent (circle in this example) and vice versa for the shape agent. This creates a situation where the agents need to exchange information in order to successfully navigate to the right landmark.\\ %This formulation ensures the agents have to learn when to and what to communicate. We expect the communication be relevant to the task and environment.\\      
\textbf{Task 2: Multi target consensus:} For this task, the map contains six landmarks each with shape and color properties and no two landmarks have the same set of properties. However, there can be two landmarks with the same color or shape and the target landmark is unique when both properties are considered. The goal is for both agents to move to the target landmark. The agent observation and action spaces are similar to the previous scenario, but here the property of the target landmark is specified in the agent's own modality. Specifying the encoding for circle, as the target, to the shape agent does not trivially solve the problem as there can be two circles and co-ordination with the color agent is necessary to figure out which circle is blue and then move towards it. In this task, the agents need to reach a consensus on which is the target landmark by learning to reasoning over their observation spaces. \\ %This is not a trivial case as now there can be two circles between which the shape agent much choose. goal vector~$g$, supplied to each agent, corresponds to its sensor modality~$m$. For example, the color-agent will have the color of the target---its corresponding encoding---in its goal vector, for each episode. The goal for both the agents is to reach the target.\\
\textbf{Task 3: Collaborative localization:} Here the setup is similar to the information exchange task. However, no target landmark is specified as the goal is for the agents to meet with each other in the shortest possible time. Here a constant negative reward $R_t$ supplied to the agents at each time-step to encourage meeting up fast. Here the agents must learn to estimate their relative position with respect to each other and then take actions to move closer.\\
Summarizing, all the above tasks share the following key challenging characteristics. Agents have (i) different sensing modalities; (ii) No knowledge about other agent's sensor or position; (iii) No common world coordinate frame in their state space; (iv) a finite communication bandwidth. \\
\textbf{Reward structure and learning framework:} We primarily used three types of rewards: $R_{d} = \sum_{i=1}^{i=n} \sqrt{x_{iT}^2+y_{iT}^2}$, where $i$ is the agent number, $x_{iT}$ and $y_{iT}$ are the horizontal and vertical distances of $i^{th}$ agent from the target landmark, and $n$ is the number of agents. We define an instantaneous reward $R_{i} = H$, where $H$ is a large number if at least one agent is touching the target at the current time step and 0 otherwise. For collaborative localization task, $R_t$ is a constant penalty per time step and $R_d$ is the inter agent distance. For all tasks, our reinforcement learning (RL) framework is based on the MADDPG algorithm (\cite{lowe2017multi}), which relies on centralized training and decentralized execution, making it suitable for multi-agent problems. The core of MADDPG is an actor-critic scheme (\cite{grondman2012survey}) that maintains a critic for each agent and the critics have access to actions (movement and communications) and rewards of all the agents. This helps with the problem of non-stationarity in multi-agent environments. In all experiments, we parameterize the output of both actors and the critics with a three layered fully connected network with ReLU activations. It must be noted that, although the agent state space allows for real numbers in the communication stream, the use of Gumbel-Softmax estimator (\cite{jang2016categorical}) transforms these into discrete valued messages. While $2^k$ word variations are possible, we observe that the agents limit their vocabulary use to $k+1$ words. For 3 channels of communication, the word vocabulary was limited to $w_0=[0,0,0], w_1=[1,0,0], w_2=[0,1,0], w_3=[0,0,1]$. Details of the learning framework, training hyperparameters and reward curves are provided in Appendix A.
%\begin{figure}
%\centering
%\includegraphics[width=0.8\textwidth]{figures/framework}
%\caption{\textit{MADDPG framework \cite{lowe2017multi} for multimodal environment}}\vspace{-10pt}
%\label{fig:1}
%\end{figure}

\section{Results}
\begin{figure}
  \begin{minipage}[c]{0.68\textwidth}
    \includegraphics[width=\textwidth]{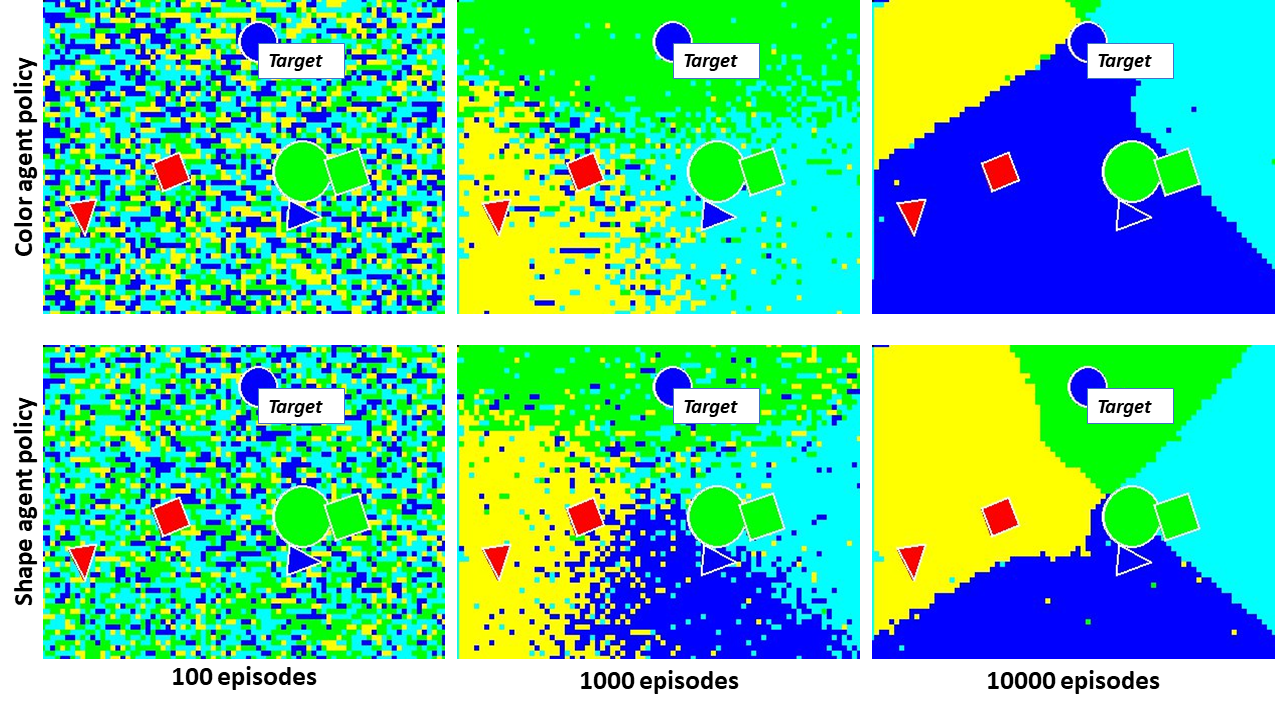}
  \end{minipage}\hfill
  \begin{minipage}[c]{0.3\textwidth}\vspace*{-2em}
    \caption{\textit{Multi target consensus task. Evolution of policies, for the word utterance $[1,0,0,0]$, for both color agent (top row) and shape-agent (bottom row). Policy relation to color: Green - Go down, Turquoise - Go left, Blue - Go up, Yellow - Go right, Grey - No movement. The meeting point of the different colors (vertex) is the equilibrium point at which the agent may come to rest.
    }} \label{fig:2}
  \end{minipage}
\end{figure}

\begin{figure}
\vspace*{-0.2in}
\centering
\includegraphics[width=0.48\textwidth]{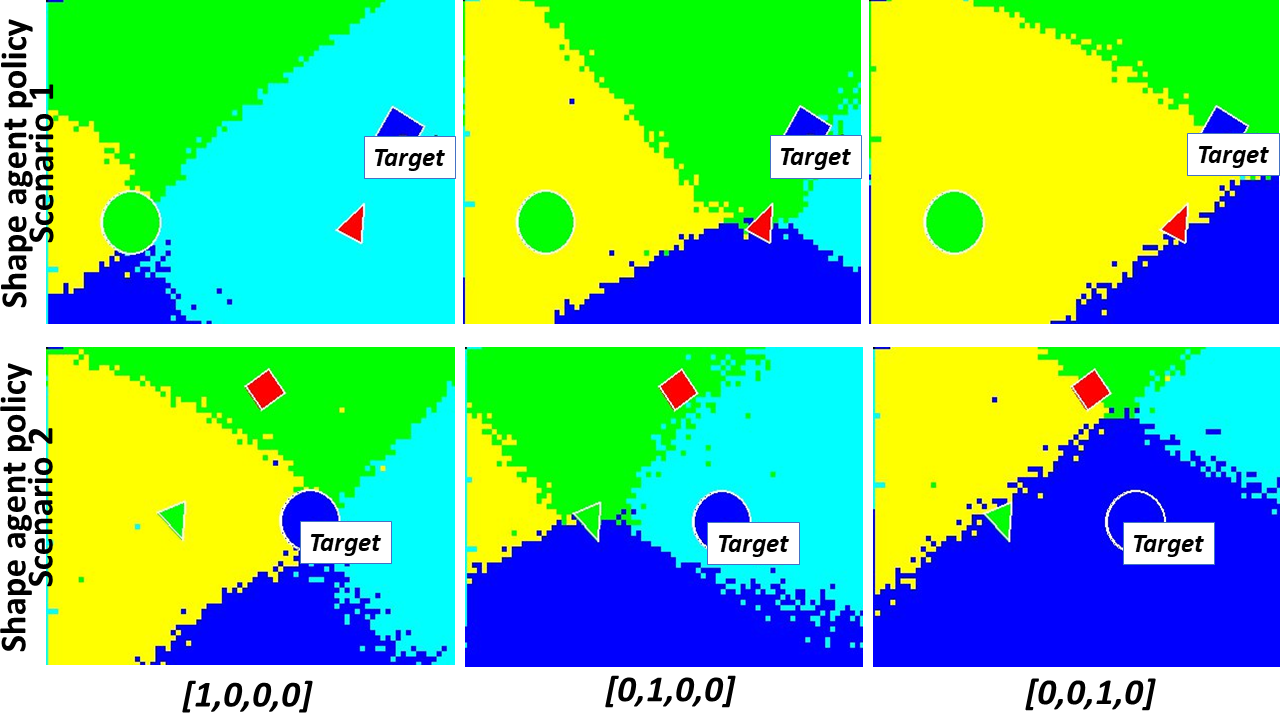}\hfill
\includegraphics[width=0.48\textwidth]{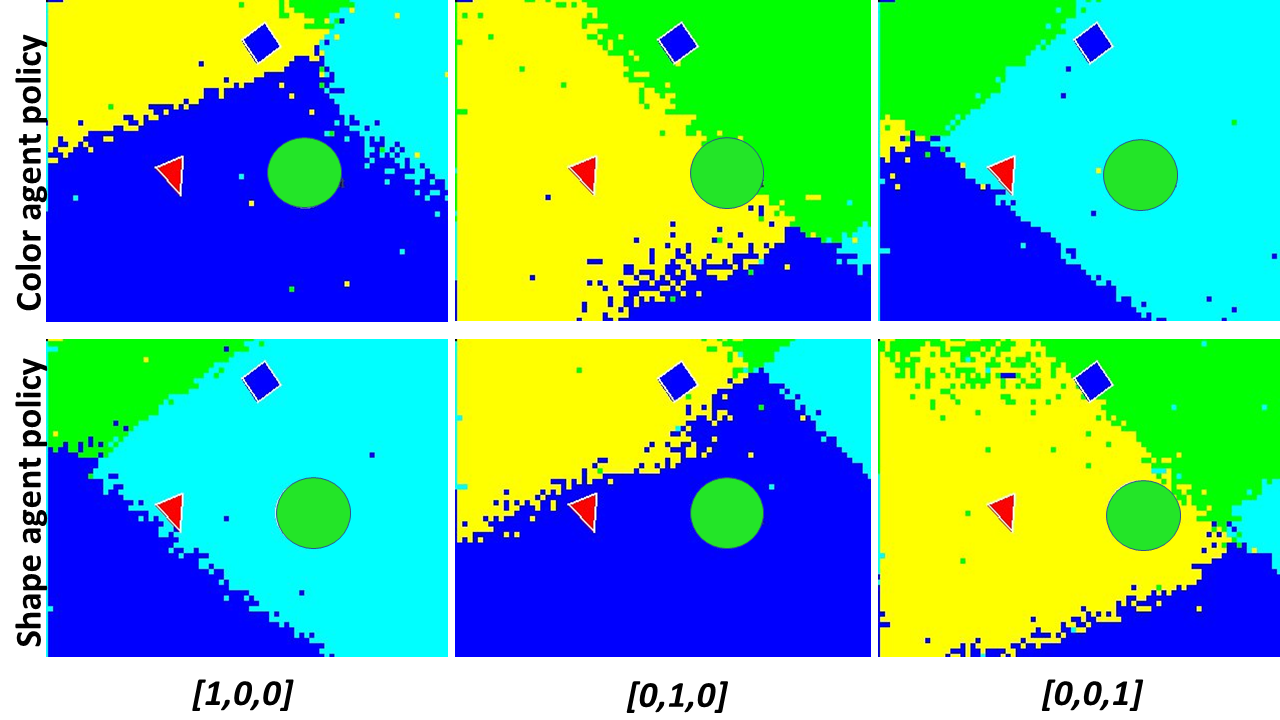}
\caption{\textbf{Left}-\textit{Information exchange task. From left to right, the color-agent utters the words $[1,0,0,0]$, $[0,1,0,0]$, $[0,0,1,0]$ to the shape agent. Top and bottom rows represent different landmark configurations.}
\textbf{Right}-\textit{Collaborative localization task. From left to right, policy for the words $[1,0,0]$, $[0,1,0]$, $[0,0,1]$ are visualized for both color (top row) and shape agents (bottom row).}}\vspace*{-0.2cm}
\label{fig:3}
\end{figure}

In the following we evaluate agent performance using simple metrics and then provide interpretations of the emerged communication between the agents.\\
\textbf{Performance metrics:} We use two metrics to evaluate the performance of the agents in achieving the common goal. Let $m_1$ denote the number of episodes in which at least one of the agent reaches the target landmark, $m_2$ denote the number of episodes where both agents reach the target landmark, and $N$ denote the total number of test episodes. We use $N=1000$, let $M_1 = m_1/N$ and $M_2 = m_2/N$. For the collaborative localization task $m_2$ denotes the number of times the agents meet with each other, and $m_1$ is the starting distance between agents divided by total distance travelled by both  agents in the episode. As we mentioned, $k$ is the number of communication bits available to the agent. Table \ref{table:1} shows these metrics for the three different tasks.\\
\begin{table}[t]
\centering
\begin{minipage}[c]{0.6\textwidth}
\begin{tabular}{|c|c|c|c|c|}
\hline
\multirow{2}{*}{\textbf{Task}} & \multirow{2}{*}{$\mathbf{k}$} & \multirow{2}{*}{\textbf{Reward}} & \multicolumn{2}{l|}{\textbf{Metric (\%)}} \\ \cline{4-5} 
 &  &  & $\mathbf{M_1}$ & $\mathbf{M_2}$ \\ \hline\hline
\multirow{3}{*}{Information exchange} & 2 & $R_{d}$ & 99 & 80 \\ \cline{2-5}
 & 3 & $R_{d}$ & 100 & 89.5 \\ \cline{2-5}
 & 4 & $R_{d}$ & 100 & 98.9 \\ \hline
\multirow{2}{*}{Multi target consensus} & 4 & $R_{d}$ & 15.1 & 0.5 \\ \cline{2-5}
 & 4 & $R_{d} + R_{i}$ & 91.8 & 61.6 \\ \hline
Collaborative localization & 3 & $R_{d} + R_{t}$ & \textit{80.5} & \textit{100} \\ \hline
\end{tabular}
\end{minipage}\hfill
%\vspace{2pt}
\begin{minipage}[c]{0.35\textwidth}
\caption{Performance metrics for the three test tasks with variations in reward structure and communication channels. \vspace*{-0.2cm}}
\end{minipage}\hfill
\label{table:1}
\vspace{-15pt}
\end{table}
\textbf{Emerged communication:} Table \ref{table:1} shows that the agents are able to successfully complete the information exchange and collaborative localization tasks almost every time. In the more complex multi target consensus task the agents achieve a $61.6\%$ success rate which is better than random chance of $25\%$. In order to better understand how the communication aid the agents, we devise a way to visualize this relation as follows. At every time step the $i^{th}$ agent decides its actions based on its observations $\mathbf{o_i}$ comprising of relative position to the landmarks, the word received from the other agent and target landmark (if applicable). For a given test case the target landmark is fixed. Then for every possible word that can be received, we can place the agent in a fixed position of the environment and query the learnt policy to find in which direction the agent would move. We can then repeat this for all possible agent positions and color code its preferential direction of motion at each location, obtaining a picture as shown in figure \ref{fig:2}. As expected, we observe random motion in the beginning of the training. Over time the color agent learns to move to the blue triangle if the word uttered by the shape agent is $[1,0,0]$. Similarly the shape agent learns to move towards the blue circle for the same utterance by the color agent. Visualization for other word utterances are given in the Appendix, see figure \ref{detailed_evo}. Note that both agents are either focusing on a blue or circular object for all word utterances as the given target in this example is a blue circle. It is remarkable that the agents are able to solve a complex map alignment and reasoning problem by directing each others actions through communication. Examples of the final learnt policy for the \emph{information exchange} and \emph{collaborative localization} tasks are shown in figure \ref{fig:3} left and right respectively. In the information exchange task, we observe that each unique word uttered by the color agent causes the shape-agent to move close to a specific shape of landmark irrespective of its current position in the map (e.g $[1,0,0,0]$ causes movement towards the circle). In the \emph{collaborative localization} task each unique word uttered by the color agent causes the shape agent to move towards a focus point in the map and vice-versa. During the episode, both agents continuously change their utterance in order to force their partner to travel towards each other and meet in the shortest time possible.\\
%Though evolution plots are presented only for the \textbf{multi target consensus} case, similar behaviors are observed for other tasks.   
%Figure \ref{fig:3} (left) shows the communication strategies learnt for the \textbf{information exchange} environment. In  Figure \ref{fig:3} (right)- \textbf{collaborative localization}, agents learn to associate words with individual landmark positions and decide on meeting close to those locations, effectively giving rise to a reference based communication protocol.
%One notable feature for trained communication policies in each of the environments is that one has the ability to focus the movement policy of another agent around the target landmark as the common vertex where the colors are divided is always close to the target location. 
%The policy map is constructed based on color encoding of the argmax of the movement policy. If the argmax of the output policy for movement for the agent is for going down, then the corresponding pixel on the map is painted green, if going left then turquoise, if go up then blue, if go right then yellow, if no movement then grey. 
%\begin{figure}
%\centering
%\includegraphics[width=1.0\textwidth]{figures/color_evo}
%\caption{\textit{Evolution of policies for color agent (Target- blue circle)}}\vspace{-10pt}
%\label{fig:3}
%\end{figure}
\textbf{Effect of communication channels:} 
Changing the number of channels affects the learning capacity. In the information exchange task when 3 channels are provided agents associate their target landmarks as ``topmost",``leftmost", or ``bottom most" landmark in the map, depending on received communication. While this is a clever reference, as the ``topmost" etc. landmark can easily be the same for the two agents as their reference axes are only translated from each other. However, this leads to a failure when the target landmark is located in the middle or is the ``rightmost". When using 4 channels agents can directly associate their targets to the property communicated to them by the other agent. In figure \ref{fig:3} (left), shape agent interprets $[1,0,0,0]$ from the color agent as a signal to go to a circle, $[0,1,0,0]$ to a triangle and $[0,0,1,0]$ to a square. This improves performance greatly.
%Changing the number of communication bits $k$ affects the nature of communication learnt, and hence the performance of the algorithm. For information exchange task, different strategies were learnt when 3 and 4 communication channels were used. In the 3 channel case, the agents learn to associate a word with a landmark position directly as in the provided examples in figure \ref{fig:3}. However, when 4 channels were used, they associated a direction to each word, similar to the multi target consensus policies shown in figure \ref{fig:2}, which achieves better performance. 
%e\begin{figure}
%\centering
%\includegraphics[width=0.6\textwidth]{figures/depth_color_evo}\hfill
%%\includegraphics[width=0.5\textwidth]{figures/depth_evo}\hfill
%%\includegraphics[width=0.5\textwidth]{figures/color_evo}
%\caption{\textit{Multi target consensus environment. Evolution of policies for shape agent (top) and color agent (bottom) (Target- blue circle). Evolution is shown for the word utterance [0,1,0,0]. Notice that different agents learn different meanings but by switching usage of words they can collaborate. Agent movement policy (pixel wise) color meanings- Green-Go down, Turquoise- Go left, Blue- Go up, Yellow- Go right.}}\vspace{-10pt}
%\label{fig:2}
%\end{figure}
%In many cases in a 2D grid world 4 prominent directions of movement were associated with 4 unique words. Therefore in all of the environments we used 4 as the default communication channel size $k$. However we also came upon this number based on experiments with different channel sizes as we found out 4 tends to work out best. 
%In the information exchange task, performance was drastically improved when the number of communication channels were increased from 3 to 4. Performance rapidly deteriorated when the communication channels were decreased below 3. Training took more time to converge with same meaning assigned to multiple words when the number of channels were increased beyond 4. 
%\\
%\textbf{Effect of changing Reward structures}
%\\
%Changing reward structures had effect on the nature of the communication learned. We primarily used three types of rewards: $R_{d} = \sum_{i=1}^{i=n} \sqrt{x_{iT}^2+y_{iT}^2}$, where $i$ is the agent number, $x_{iT}$ and $y_{iT}$ are the horizontal and vertical distance of $i^{th}$ agent from the target landmark, and $n$ is the number of agents, $R_{i} = H$. $H$ is a large instantaneous reward if at least one agent is touching the target at the current time step and 0 otherwise. For collaborative localization task, $R_t$ is a constant penalty per time step and $R_d$ is the inter agent distance. More discussion is provided in supplementary material.
%\\
\section{Conclusion}
We studied the application of multi agent reinforcement learning for task driven multimodal decision making. We analyzed the emergence of interpretable communication between agents and found that adaptive and non trivial communication protocols can be learned based on number of available communication channels and imposed reward structures.  The size of the communication channel can be crucial in deciding the amount of information that is required to reconcile various modalities with each other and reward structures affect the nature of the learned communication. Visualizing emergent policies in the agent's action spaces confirms that powerful joint representation of the world can be encoded through communication.

\section{Acknowledgement}
We are thankful to Shashank Shivkumar at Honeywell Aerospace for discussions related to this paper ranging from initial idea for the research through algorithm implementation and testing. 
%\section*{References}
\bibliographystyle{plainnat} % or try abbrvnat or unsrtnat or plainnat
\bibliography{biblio}

\newpage
\setcounter{page}{1}
\begin{center}
\Large{\textbf{Appendix: A perspective on multi-agent communication for information fusion}}
\end{center}

\section*{A : Framework and training}
\begin{figure}[htp]
\centering
\includegraphics[width=0.98\textwidth]{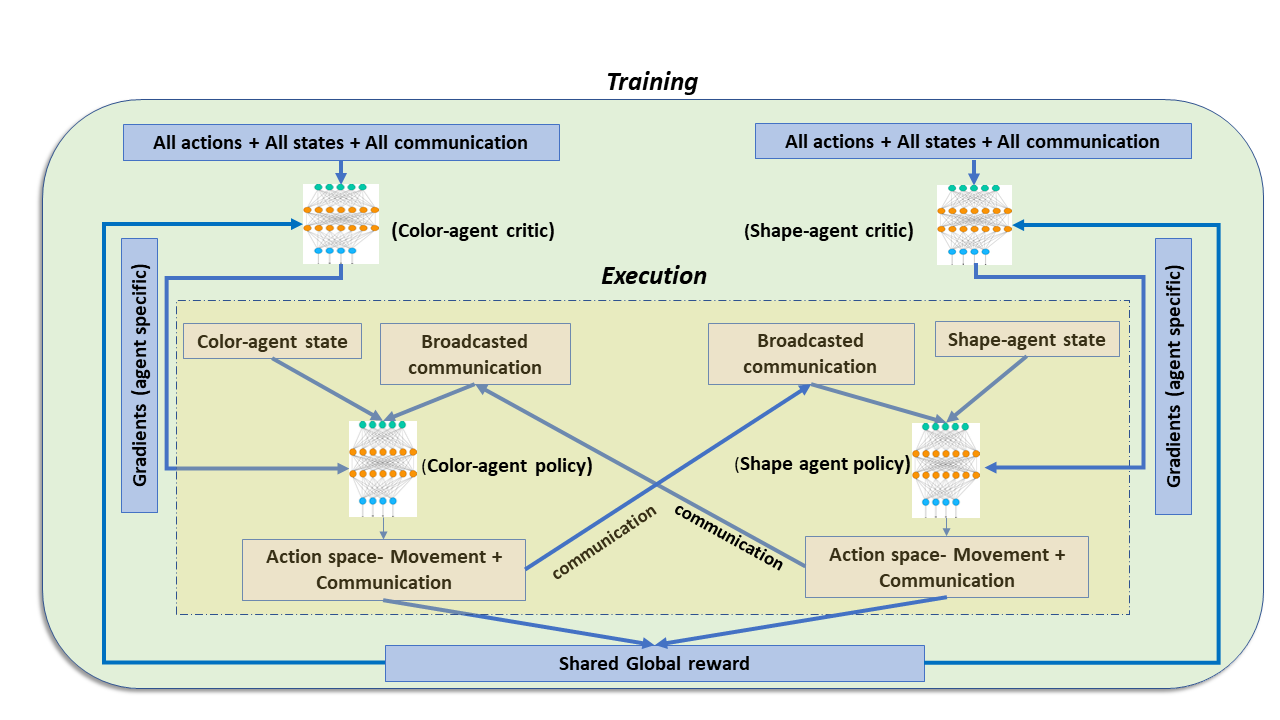} %or .PDF or .EPS
\caption{\textit{Learning framework used in this work.}}\vspace{-10pt}
\label{framework}
\end{figure}

\begin{figure}
\centering
\includegraphics[width=1.0\textwidth]{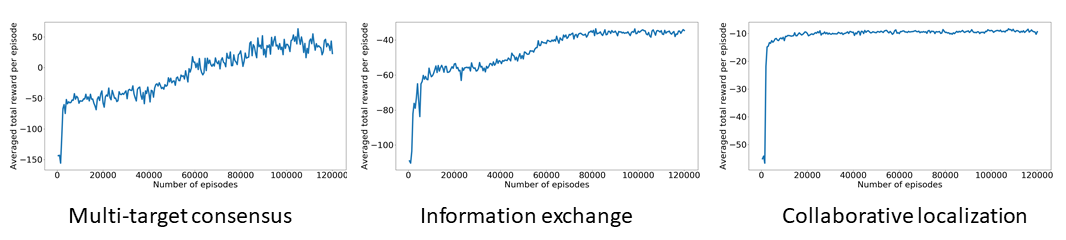}
\caption{Plots of average total cumulative rewards vs training episode number \textit{Left- multi target consensus task. Middle- information exchange task. Right - collaborative localization task.}}\vspace{-10pt}
\label{training_prog}
\end{figure}

An overview of the learning framework we used for the three different tasks is shown in Figure \ref{framework}. In the information exchange and collaborative localization tasks we train with a maximum episode length of 60. In multi-target consensus task, agents are trained for 120000 episodes with a maximum episode length of 80 steps. We use Adam optimizer with learning rate of 0.01, discount factor of 0.001 for the critics, and a batch size of 1024. In general, convergence was observed within 5000 episodes. 
Figure \ref{training_prog} shows the training progress over number of episodes for the three different tasks. In the information exchange and collaborative localization tasks, the majority of the policy improvement takes place in the initial episodes and the agent maintains the word-action associations it learns over the following episodes. In the more complex multi target consensus task, the agents take the longest to learn meaningful policies that maximize reward. Reward curve has a sudden peak near 5000 episodes, however the policies still gradually keep improving over time till 80000 episodes. 

\section*{B : Effect of reward structure on communication}
\begin{figure}
\centering
\includegraphics[width=0.95\textwidth]{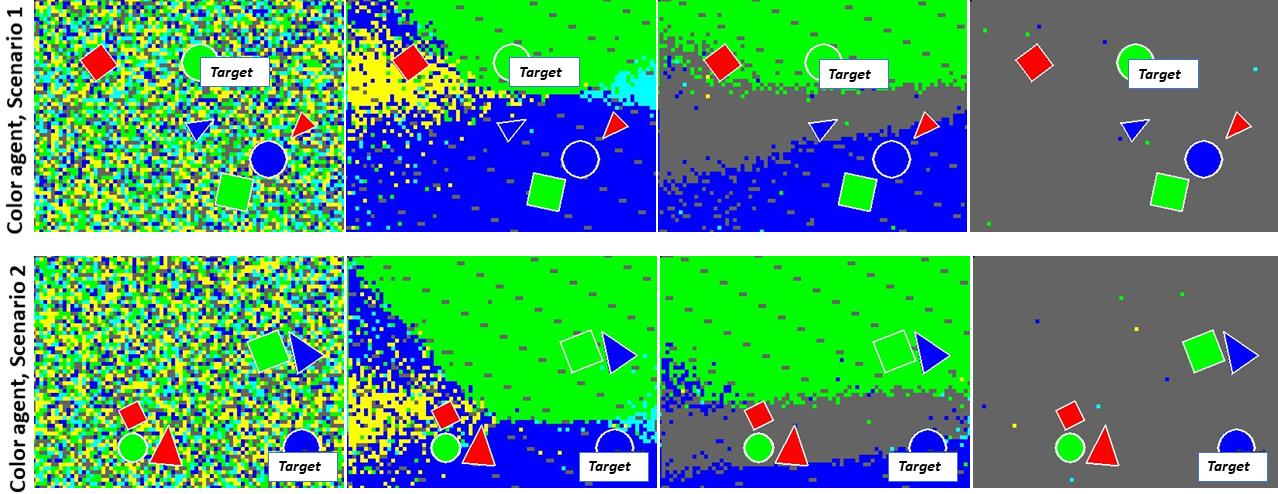}
\caption{\textit{Evolution of policy in multi target consensus when only average distance penalty is used. Two scenarios are shown. Identical plots are obtained for shape agent.}}\vspace{-10pt}
\label{policy_erode}
\end{figure}

%When MADDPG (\cite{lowe2017multi}) was introduced, the authors mainly focused on a team reward assignment based on the average of the distances of all the agents from the target. It can be described as $R_{d} = \sum_{i=1}^{i=n} \|x_i\|$, where $i$ is the agent number and  $n$ is the number of agents. However, 
For the complex multi-target consensus task, we found that just using a continuous average distance penalty ($R_{d}$) for each time step in the episode is detrimental to learning. As the training progresses, the agents learn to ignore communications and stay in the same place where they started: the whole policy map changes to grey (no movement) in the final epochs as shown in figure \ref{policy_erode}. To encourage more exploration, we introduced an instantaneous touching reward $R_{i}$ in addition to the constant average distance penalty and observe improved performance. We also experimented with just using $R_{i}$ alone and the policies learned are sub optimal. So it appears that both the reward types are necessary for learning in complex multimodal scenarios.

%It can be described as $R_{i} = H$ if at least one agent is touching the target at the current time step and 0 otherwise. $H$ is generally greater than 1. On using instantaneous touching reward alone however, the policies learned are sub optimal. So it is clear that both the reward types are necessary for learning in multimodal scenarios.

%It is clear that just using constant distance penalty $R_d$ in multi target consensus is not enough to achieve optimal policy. Figure \ref{policy_erode} shows the evolution of communication for the case when only $R_d$ is used. It can be seen that over time agents learn to ignore each others communication and stay in the same place where they are started as the whole policy visualization turns grey (no movement no matter where the agent is in the map).

\begin{figure}
\centering
\includegraphics[width=0.8\textwidth]{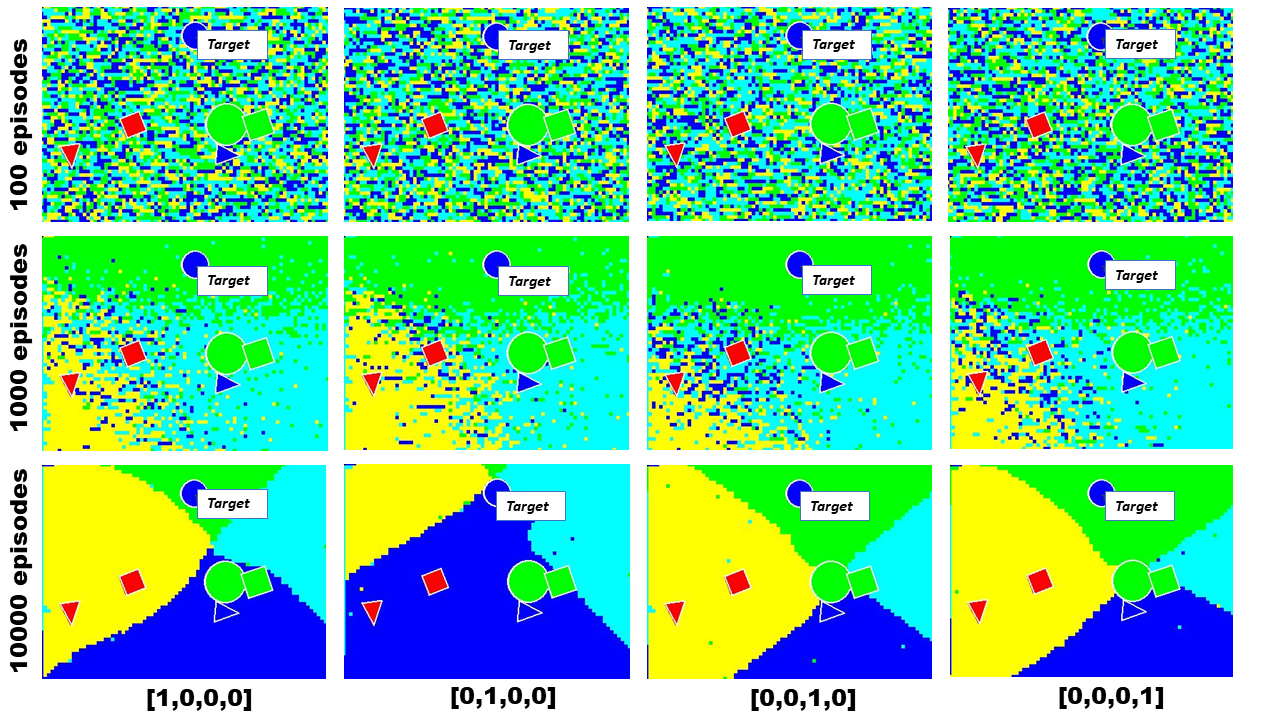}
\hfill
\includegraphics[width=0.8\textwidth]{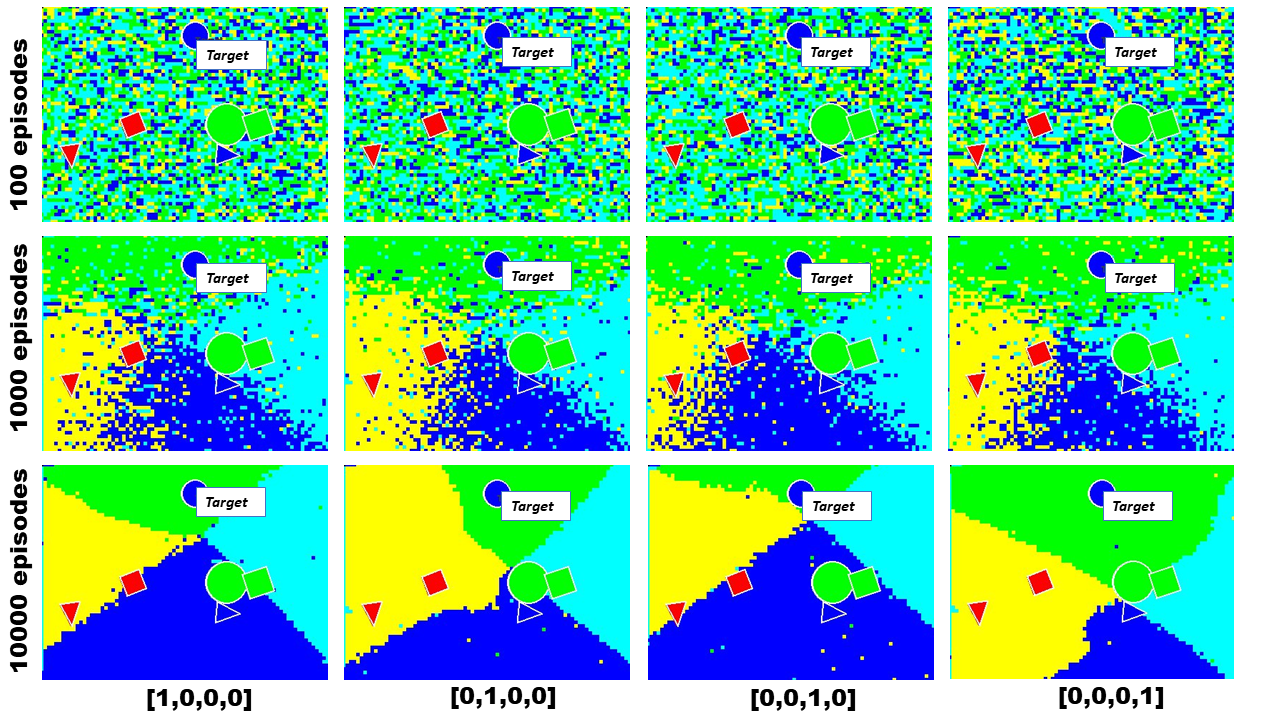}
\caption{\textit{Evolution of policy in multi target consensus for shape-agent (top) and color-agent (bottom) for different words uttered (left to right).}}\vspace{-10pt}
\label{detailed_evo}
\end{figure}

\section*{C : Number of words learned}
While $k$ bit communication channel is used, we experimentally verified that $2^k$ different words are valid. However the agents mainly chose to use only $k+1$ words while navigating. An example policy evolution corresponding to all the words in multi-target consensus is shown in figure \ref{detailed_evo}.

\begin{figure}
\centering
\includegraphics[width=0.95\textwidth]{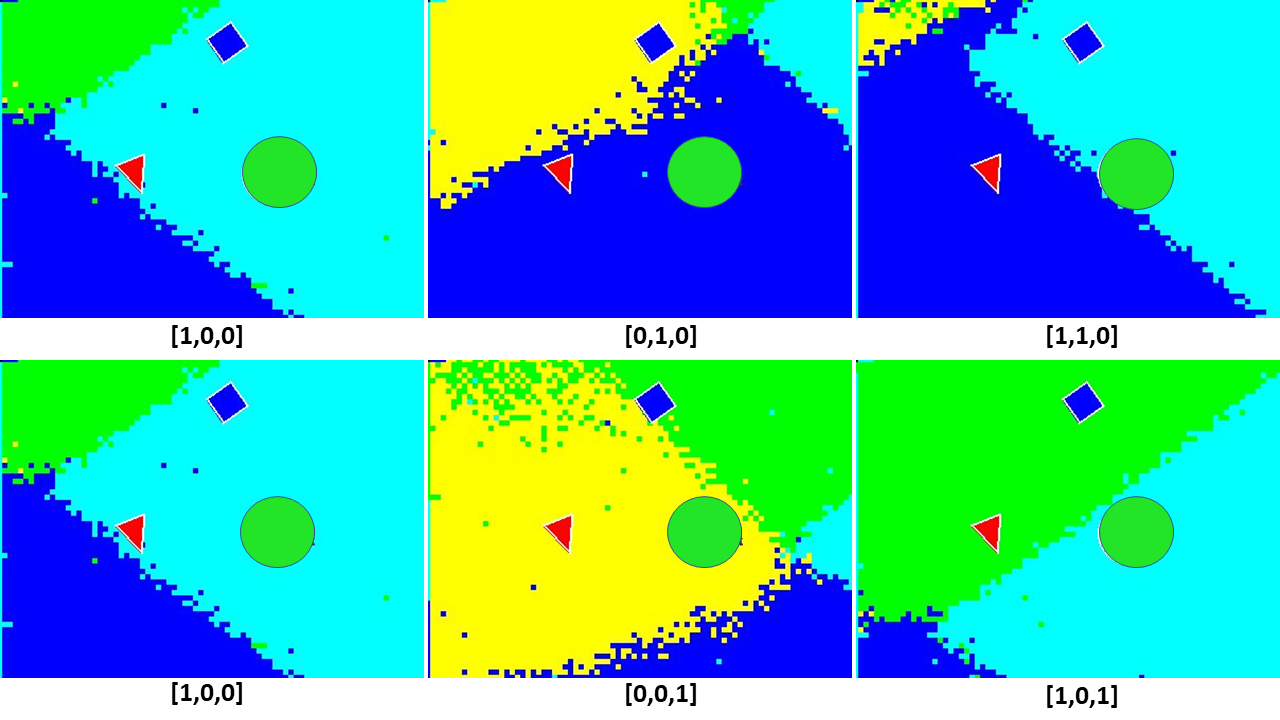}
\caption{\textit{Policy visualizations corresponding to simple $[1,0,0],[0,1,0],[0,0,1]$ and complex words $[1,1,0],[1,0,1]$ in collaborative localization task.}}\vspace{-10pt}
\label{composition}
\end{figure}

To understand what the other unused words meant we visualized the policy corresponding to two such words as shown in figure \ref{composition}. We find that the (combination) word utterance $[1,1,0]$ will produce a new focus / equilibrium point different from that of the individual words $[1,0,0]$ and $[0,1,0]$. While it can be useful to use another focus point to guide the movement of the other agent in the collaborative localization environment, the agents prefer to just utter different simple words sequentially, rather than a more complex word once.

\end{document}